\documentclass[manuscript]{aastex}

\usepackage{natbib}

\shorttitle{Very metal-poor Cepheid models}
\shortauthors{Marconi et al.}

\begin{document}


\title{Pulsation Models for Ultra-Low ($Z=0.0004$) Metallicity Classical Cepheids}

\author{M. Marconi \altaffilmark{1}, I. Musella \altaffilmark{1}, G. Fiorentino \altaffilmark{2,3}, G. Clementini \altaffilmark{2}, A. Aloisi \altaffilmark{4},
F. Annibali \altaffilmark{5}, R. Contreras Ramos \altaffilmark{2}, A. Saha\altaffilmark{6}, M. Tosi \altaffilmark{2}, R. P. van der Marel\altaffilmark{4}}

\altaffiltext{1}{INAF $-$ Osservatorio Astronomico di Capodimonte, Via Moiariello 16, 
80131 Napoli, Italy; marcella@na.astro.it; ilaria@na.astro.it}

\altaffiltext{2}{INAF $-$ Osservatorio Astronomico di Bologna, Via Ranzani 1, 40127 Bologna, Italy; 
gisella.clementini@oabo.inaf.it, 
rodrigo.contreras@oabo.inaf.it; monica.tosi@oabo.inaf.it}
 
\altaffiltext{3}{Kapteyn Astronomical Institute, University of Groningen, Postbus 800, 9700 AV Groningen, The Netherlands; fiorentino@astro.rug.nl}

\altaffiltext{4}{Space Telescope Science Institute, 3700 San Martin 
Drive, Baltimore, MD 21218; aloisi@stsci.edu} 

\altaffiltext{5}{INAF $-$ Osservatorio Astronomico di Padova;
francesca.annibali@oapd.inaf.it}

\altaffiltext{6}{National Optical Astronomy Observatory, P.O. Box 26732, Tucson, AZ 85726}

\begin{abstract}
Classical Cepheids are primary distance indicators playing a
fundamental role in the calibration of the extragalactic distance
scale. The possible dependence of their characteristic
Period-Luminosity (PL) relation on chemical composition is still
debated in the literature, and the behaviour of these pulsators at
very low metallicity regimes is almost unexplored. In order to derive
constraints on the application of the Period-Luminosity relation at
low metal abundances, we investigate the properties of the few
ultra-low metallicity ($Z \approx 0.0004$) Cepheids recently
discovered in the Blue Compact Dwarf galaxy IZw18.  To this purpose we
have computed an updated and extended set of nonlinear convective
models for $Z=0.0004$ and $Y=0.24$, spanning a wide range of stellar
masses, and taking into account the evolutionary constraints for
selected luminosity levels.  As a result we are able to predict the
topology of the instability strip, the variations of all the relevant
quantities along the pulsation cycle, including the morphology of the
light curves, the theoretical Period-Luminosity-Color,
Period-Wesenheit and Period-Luminosity relations at such a low
metallicity. For each of these relations we provide the appropriate
coefficients for fundamental mode pulsators with Z=0.0004.  By
comparing these results with the properties of more metal rich
Cepheids we find that the synthetic PL relations for $Z=0.0004$ are
steeper than at higher Z, but similar to the $Z=0.004$ ones, thus
suggesting a leveling off of the metallicity effect towards the lowest
Zs.
\end{abstract}  



\section{Introduction}
With their characteristic Period-Luminosity (PL) and
Period-Luminosity-Color (PLC) relations, Classical Cepheids play a
relevant role as distance indicators for the definition of the
extragalactic distance scale.  The application of a Large Magellanic
Cloud (LMC)-based PL relation to external galaxies observed with the
Hubble Space Telescope (HST), has led to the calibration of secondary
distance indicators and in turn to an estimate of the Hubble constant
\citep[$H_0$, see e.g.][]{fre01,sa01}.
 
The theoretical explanation for the observational evidence of a PL 
relation for Classical Cepheids relies on the
assumption that intermediate mass stars undergoing central Helium
burning are characterized by a Mass-Luminosity (ML) relation, as
predicted by stellar evolution models.
 
The ML relation is significantly dependent on several physical and
numerical assumptions adopted in stellar evolution models, thus the
comparison between observations and Cepheid evolutionary/pulsation
models provides a unique insight into the stellar evolution and
pulsation physics.

An important issue still under debate is the universality of the
Cepheid PL relations, and, in turn, the possibility of applying
LMC-calibrated Cepheid PL relations to infer the distance of any
galaxy containing Cepheids, independently of its chemical composition.
In the last decade many efforts were devoted to investigate the
dependence of the Cepheid properties on metallicity, since a
metallicity effect could produce significant systematic errors in the
evaluation of the extragalactic distance scale and in turn on $H_0$. 
However, no general consensus has been reached yet.

On the theoretical side, linear non adiabatic models mostly suggest
that a variation in chemical composition produces negligible effects
on the PL relations \citep[see e.g.][]{cwc93,a99,sg98,san99}, while
nonlinear convective pulsation models \citep{bms99,f02,mmf05} 
predict a significant metallicity effect on the Cepheid instability
strip (hereinafter IS) topology and on the PL relations.  For instance,
\citet{cmm00} and \citet{f02} find that the synthetic PL
relations become shallower as the metallicity increases, and the size of
 the effect depends
also on the photometric band, with a decreasing sensitivity
as wavelength increases from optical to near infrared (NIR)
bands. Consequently, metal-rich pulsators with periods longer than
five days have fainter optical magnitudes than metal-poor
pulsators. This scenario is further complicated by the theoretical
finding that the helium mass fraction $Y$ also plays a role at the highest
metallicities ($Z \ge 0.02$), with the slope of the linear PLs
increasing as $Y$ increases at fixed $Z$ \citep{f02,mmf05}. In turn,
the metallicity correction to the predicted distance moduli varies
with the assumed $\Delta{Y}/\Delta{Z}$, showing a sort of turnover
around solar metallicity.  In particular, for P $\ge$ 20 days and
[O/H] $\ge$ 0.2 dex, as measured in several spiral galaxies observed
by the HST Key Project \citep{fre01}, the average predicted
metallicity correction varies from $\sim -$0.2 mag to $\sim$ +0.25 mag, 
as $\Delta{Y}/\Delta{Z}$ increases from 2.0 to 3.5.

On the empirical side, several authors suggest that metal-rich
Cepheids are, at fixed period, brighter than metal-poor ones, either
over the entire period range \citep[][]{KSS98,K03,SCG04,G04,S04,M06},
or at least for periods shorter than $\approx$ 25 days
\citep[][]{SBR97,san04}. \citet{S04} by comparing
distances based on Cepheids and on the Tip of the Red Giant Branch
(TRGB) method for a selected sample of spiral galaxies, found that the
metallicity effect, usually expressed in the form
$\gamma=\delta{\mu_0}/\delta{\log{Z}}$, is equal to $-$0.25 mag
dex$^{-1}$. However, the revised TRGB distances by
\citet{r07} no longer support a $\gamma$ value of $-$0.25 mag
dex$^{-1}$ and are in better agreement with the theoretical
results \citep[see][for details]{b08}.  Other empirical results that
seem to support the nonlinear theoretical scenario were obtained by
\citet{r05,r08}, on the basis of spectroscopic [Fe/H] measurements for
Galactic and Magellanic Cloud Cepheids. Using only their spectroscopic
abundances for Galactic Cepheids with published distances, \citet{r05}
found that the metallicity correction is in qualitative agreement with
the model results, and shows the same kind of turnover around solar
metallicity as predicted by nonlinear pulsation theory.  More recently, using direct high
resolution metallicity measurements for a total of 68 Galactic and
Magellanic Cepheids, \citet{r08} found that metallicity affects the
$V$-band Cepheid PL relation, with metal-rich Cepheids appearing to be
systematically fainter than metal-poor ones, in agreement with the
theoretical predictions by \citet{b99} and \citet{cmm00}, and
independently of the adopted distance scale for Galactic and LMC Cepheids.

On the other hand, direct empirical tests of the metallicity effect
based on the observation of Cepheids in the outer and inner fields of
M101 \citep{KSS98}, in NGC4258 \citep{M06}, and, recently, in M33
\citep{m33}, should be taken with caution. Indeed, in M101 the
presence of blended Cepheids could have affected the results of the
test \citep[see][]{M06}, while in the case of NGC4258 there is
evidence against the assumed metallicity gradient, as both the
comparison with pulsation models and the most recent HII abundance
measurements \citep{d00} suggest a rather constant LMC-like
metal-content for the Cepheids observed in the two selected fields
\citep[see][for details]{b08}. In M33, \citet{m33} adopted the [O/H]
gradient by \citet{Mag07} and conclude that blending cannot
account for the difference in distance modulus between the two
fields. However, as also pointed out by \citet{r08}, these results
are always based on indirect measurements of the metallicity, assumed
to be that corresponding to the oxygen nebular abundances of HII
regions at the same Galactocentric distance of the Cepheids.
  
  Finally, there are also recent empirical results in favour of the
  universality of the PL slope: distance estimates based on the
  ``near-infrared surface brightness'' (ISB) technique \citep[][and
  references therein]{f07} and the HST parallaxes \citep{Ben07} for
  ten Galactic Cepheids seem to indicate a vanishing metallicity
  effect between Galactic and Magellanic Cepheids.  Similar
  suggestions have been put forward by several authors \citep[see
  e.g.][]{pie06,pie07,gie08} for Cepheids in galaxies with
  metallicities significantly lower than the LMC, but no theoretical
  predictions are available in the literature for the metallicity
  effect on the properties of Cepheids in this range of abundances or
  at lower Z.\par

The identification and study of Cepheids in very metal-poor galaxies,
is a challenging perspective. These galaxies represent the
closest analog to primordial galaxies in the early universe and, as
such, offer the best place where to study star formation and stellar
evolution in a (almost) pristine environment. Some of these galaxies have also 
been claimed to be
primordial systems just forming in the local Universe due to lack of
detection of faint red stars on the red giant branch tip (TRGB, stars
older than $\sim$1 Gyr). However, the lack of these stars may also be due to
a much larger distance of these systems that does not allow to resolve
the faint TRGB stars. A precise determination of the distance of
these galaxies through Cepheids allows us to verify or confute their
nature of young nearby galaxies.  In the context of the HST ACS GO program
10586 (PI: A. Aloisi) we have identified three Classical Cepheids in
the Blue Compact Dwarf galaxy IZw18, having respectively periods of
8.71, 125 and 130.3 days.  A new estimate of the distance, and, in
turn, of the age of the galaxy stellar populations were obtained both
from the Classical Cepheids and from the galaxy's TRGB, which allowed
us to rule out the possibility that IZw18 is a truly primordial
galaxy of recent formation in the local universe \citep{a07}. At the
same time this project also provided a first sample of ultra-low
metallicity ($\sim 1/50 Z_{\odot}$) Classical Cepheids \citep{f10},
thus representing an important benchmark for the pulsation models.

In this paper we investigate theoretically the pulsation properties of
Cepheids at this very low metallicity (Z=0.0004), in order to probe
the metallicity effect at Z $<$ 0.004, and to provide a theoretical
scenario for the Classical Cepheids recently detected in IZw18
\citep{a07,f10}. The organization of the paper is as follows: in
Sect. 2 we present the new set of pulsation models; in Sections 3 and
4 we show our results for the topology of the IS and the behaviour of
the light curves, while the multi-filter PLC, Period-Wesenheit and PL
relations are presented and discussed in Sect. 5.  Finally, in Sect. 6
we discuss the comparison with the Cepheids observed in IZw18 and the
implications for the Cepheid distance scale. Conclusions are
summarized in Sect. 7.

\section{Pulsation models for ultra-low metallicity Cepheids}

Using the same pulsation code \citep[see][for details]{bms99} adopted
in our previous investigations of more metallic Cepheids
\citep{bms99,f02,mmf05}, we investigated for the first time the full
amplitude behaviour of metal-poor Cepheid models.  To this purpose we
fixed the metal and helium abundances at $Z=0.0004$, $Y=0.24$ and
selected model stellar masses in the range from 4 to 13 $M_{\odot}$.
Two luminosity levels (see Table 1) were assumed for most of the
selected stellar masses corresponding to canonical and mild
overshooting (noncanonical) evolutionary predictions, respectively.  The canonical
scenario is provided by the Pisa Evolutionary Library (PEL) for
  $M\le 11 M_{\odot}$ and discussed in \citet{c03,ca04}\footnote{
      For models at 13 $M_{\odot}$ we extrapolated the ML relation
      obtained for lower masses taking into account the evidence that,
      for higher metal content, models with 13 $M_{\odot}$ follow the
      same ML relation as lower mass models.}, while the noncanonical luminosity level is obtained by adding 0.25 dex (in $\log
L/L_{\odot}$) to the canonical value, according to the prescriptions
by \citet{cwc93}. For each mass and luminosity, an extensive range of
effective temperatures was explored for pulsation instability, in
order to derive the location of the blue and red edges of the IS in
the Hertzsprung-Russell (HR) diagram.  For canonical models, we also explored the effect of
varying the mixing length parameter $l/H_p$, adopted in the code to
close the nonlinear system of dynamical and convective equations
\citep[see][for details]{bs94,bms99}, in the range from the standard
value of 1.5 to 2.0.  In the following two sections we discuss the
results obtained for the topology of the IS, and for the morphological
properties of the bolometric light curves, respectively.

\section{Predicted IS for Cepheids with Z=0.0004}

For each selected mass and luminosity, the effective temperatures of
the hottest pulsating fundamental model, increased by 50 K
(fundamental blue edge, hereafter FBE), and of the coolest pulsating
fundamental model, decreased by 50 K (fundamental red edge, hereafter
FRE) are reported in the third and fourth columns of Table 1,
respectively.
The adopted uncertainty of 50 K in the location of the IS boundaries 
stems from our adopted effective temperature step of 100 K in the model 
selection. This is the procedure we adopted in our previous theoretical papers 
on Cepheids \citep[e.g.]{bms99,f02,mmf05,fio07} and RR Lyrae stars \citep{b97,m03,dc04}.

  For the first overtone mode, canonical pulsators
show a stable oscillation only for stellar masses smaller than or
equal to 7 $M_{\odot}$, while we verified that the first overtone
IS disappears increasing the mass by only 0.1
$M_{\odot}$.  Noncanonical first overtone models present a stable
limit cycle only for masses $\le 4.5 M_{\odot}$. The first
overtone IS boundaries are reported in Table 2.

Our theoretical results confirm that the longest period for
first overtone pulsators (${P^{FO}}_{MAX}$) increases as the metal
content decreases, being of about 9 days at Z=0.0004 and of about 6
days at Z=0.004 \citep{b01,b02}.  As already discussed by \citet{b02},
this behaviour is in marginal agreement with observational evidence
in the Milky Way and in the Magellanic Clouds, and at
variance with results from linear pulsation models that predict the
opposite trend.

The location in the HR diagram of fundamental and first overtone
IS boundaries is shown in Figure 1, as obtained from
canonical (middle panel) and noncanonical (upper panel) pulsation
models, respectively.  In this plot the solid lines represent the
FBE and FRE, while the dashed lines are the
boundaries of the first overtone instability strip.  We notice that at
fixed luminosity level, there is no significant difference in the
effective temperature of the IS boundaries, as we move
from a canonical to a noncanonical mass-to-luminosity (ML) relation.
For canonical fundamental models, the bottom panel of Figure 1 shows the
effect of a different assumption on the efficiency of convection,
namely $l/H_p=2.0$ (dotted lines) instead of $l/H_p=1.5$. We notice
that increasing the mixing-length parameter, the instability strip
becomes narrower, as already found in our previous investigations of
more metallic Cepheid models \citep{fio07} and for RR Lyrae stars
\citep{dc04}.

\subsection{Comparison with the predictions at higher metallicities}

Figure 2 shows the comparison between the IS obtained in the present
paper for fundamental canonical models with $l/H_p=1.5$, and those
obtained by \citet{bms99} at metallicities traditionally assumed for
the Magellanic Clouds and the Galactic Cepheids, namely $Z=0.004$,
$Z=0.008$, and $Z=0.02$.  We notice that the location of the new
boundaries confirms the trend already found at higher metal content up
to $Z=0.03$ \citep[see][]{f02, mmf05} that the predicted IS moves toward lower effective temperatures as the metallicity
increases. In fact, the FBE for $Z=0.0004$ is bluer than the one for
$Z=0.004$ because of the increased abundance of Hydrogen that produces
an higher efficiency of the Hydrogen ionization region in driving
pulsation. For the same reason, the FRE for $Z=0.0004$ is cooler than
the one for $Z=0.004$ for masses lower than about 9$M_{\odot}$. At
higher masses and luminosities (lower densities), the simultaneous
effect of the lower metallicity, combined with the high efficiency of
convection (that reduces the driving role of both Hydrogen and Helium
ionization regions),  prevails and produces a hotter red edge.

However, the effect seems to be relatively small, so that no relevant
implication on the zero point and on the slope of the theoretical PL
relations are expected when moving from the typical metal contents of
the  Magellanic Clouds to the significantly lower abundance of IZw18
(see below).

\section{An atlas of theoretical light  curves for ultra-low metallicity Cepheids}

The bolometric light curves we have obtained for canonical fundamental
pulsation models are shown in Figs. 3-8 for stellar masses ranging
from 4 to 13 $M_{\odot}$. The light curve data for the complete set of
models are available upon request from the authors.  For stellar
masses ranging from 5 to 11 $M_{\odot}$ the predicted light curves for
$l/H_p=2.0$ are also shown for comparison. Confirming our previous
results on the effect of the convective efficiency on the pulsation
properties \citep{fio07,dc04}, the amplitude of the theoretical light
curves decreases as $l/H_p$ increases from 1.5 to 2.0.  An interesting
feature shown by these models is the behaviour of the bump-Cepheids.
As well known, Classical Cepheids in the period range 6 $<$ P $<$16
days show a secondary maximum (bump) along both light and  radial
velocity curves. The Hertzsprung progression
\citep[HP;][]{h26,lw58,bms00} is the relationship linking the phase of
the bump and the pulsation period.  Inspection of Figure 5 suggests that
at Z=0.0004 the transition from light curves with the bump on the
decreasing branch to light curves with the bump on the increasing
branch occurs at a pulsation period of $\approx$ 13 days. This
value is significantly longer than the observed \citep{m92,m00,be98}
and predicted periods \citep{bms00} at the center of the HP ($P_{HP}$) for Galactic ($\approx$ 10 d), LMC ($\approx$
10.5 d) and SMC ($\approx$ 11 d) bump Cepheids. This theoretical
result thus confirms the predicted and observed trend of $P_{HP}$ with
metallicity \citep[see also][]{mmf05}.

\section{Fundamental relations for Z=0.0004 Cepheids}

The combination of the period-density relation for pulsating stars and
the Stephan-Boltzmann law provides a strict relation between the
pulsation period, the luminosity, the effective temperature and the
mass. In Tables 3 and 4, we report the stellar masses, the luminosities, the 
effective temperatures, the periods and the predicted intensity-averaged mean
magnitudes and colors (see below) for fundamental and first overtone models, respectively. 
A linear regression fit to the data reported in the first four columns
of these tables allows us to derive the following theoretical
pulsation relations:
$$\log P = 10.754 + 0.921 \log L -0.770 \log M -3.336 \log T_e  $$
for fundamental models, and
$$\log P = 10.807 + 0.818 \log L -0.618 \log M -3.324 \log T_e  $$
for first overtone models, both showing an intrinsic dispersion of
0.01 mag.  The bolometric light curves of the new models were
transformed into the observational Johnson-Cousins bands (UBVRIJK) by
means of the {\it atmosphere models} by \citet{c97a,c97b} to
derive the intensity-averaged mean magnitudes and colors reported in
Tables 3 and 4. Magnitude-averaged values were also computed and are
available upon request from the authors. Static magnitude values have
been derived and used to obtain the boundaries of the instability
strip, at each chemical composition, in the various period-magnitude
planes and, in turn, to construct synthetic multi band PL relations for
fundamental canonical models (see Table 7).

Linear regressions through the periods, magnitudes and colors
reported in Table 3 allow us to derive theoretical PLC and Wesenheit
relations for fundamental Cepheids, both in the canonical case and in
the noncanonical one. Since the two cases differ only by a luminosity
offset (0.25 dex), the relations are the same in the two cases except for the
value of the term $\log{L/L_C}$, where $L/L_C$ is the ratio between the 
luminosity and the canonical luminosity level for the corresponding chemical 
composition. The PLC has the form:

$<M_V>$ = $\alpha$+$\beta$log$P$+$\gamma$(color)+$\delta${$\log{L/L_C}$},

\noindent
where $\alpha$, $\beta$, $\gamma$ and $\delta$ have the values listed in 
Table 5 for each color. The Wesenheit relation has the form: 

$<M_V>$ -- $\gamma$(color) = $\alpha_W$+$\beta_W$log$P$+$\delta_W${$\log{L/L_C}$},

\noindent
where $\alpha_W$, $\beta_W$, and $\delta_W$ have the values listed in 
Table 6.

The coefficients for the corresponding  relations for first overtone
pulsators are available upon request.  Figures 9 and 10 show the
comparison between models at Z=0.0004 (IZw18) and Z=0.008 (LMC) in the
PLC and Wesenheit planes, respectively, when the canonical ML relation
is assumed. The magnitude differences at fixed period are mainly due
to the effect of the  different chemical composition on the predicted
luminosity of each stellar mass.   No relevant dependence on the
assumed $l/H_p$ parameter is expected. The PLC does not depend on the instability strip topology and the Wesenheit relation has only a weak dependence, thanks to the inclusion of a color term with a coefficient given by the ratio between the total and the relative extinction and thus generally different from that of the PLC \citep[see also][]{fio07}.

Following the same procedure adopted in our previous papers
\citep{cmm00,f02,mmf05}, we have used the information on the
boundaries of the IS and the pulsation relation for
fundamental models to derive synthetic PL relations in the BVRIJK
filters \citep[see][for details]{cmm00}:

$M_j= A_c + B_c \log P$, 

\noindent
where $j$ is the photometric band.  The coefficients $A_c$ and $B_c$ of the
linear relations are reported in Tables 7 and 8 for canonical models with
$l/H_p=1.5$ and 2.0, respectively, and in Table 9 for noncanonical models with
$l/H_p=1.5.$  ($A_{nc}$ and $B_{nc}$).
Quadratic relations are also available, upon
  request from the authors. However, as discussed in the text, the
  nonlinear effect is very small at the chemical composition of 
  IZw18's Cepheids.
  
 Figure 11 shows the $M_{\lambda}$-$\log P$
distribution of canonical fundamental pulsators for the two different
assumptions on $l/H_p$ (left and right panels) and for the
noncanonical models with $l/H_p=1.5$ (middle panel). The solid lines
represent the linear fits from Table 7. As already found for higher
metal contents, the distribution of the pulsators becomes narrower
when moving towards the infrared. We also find that the coefficients of the PL relations depend on
both the ML relation and the value of the mixing length parameter (as also
found from the comparison of Table 7 with Tables 8 and 9), but with an
effect on the inferred distance lower than 10 \%.
 
The PL relations at Z=0.0004 are remarkably more linear than those
obtained at higher Z with no clear evidence of a break at 10 days (see \citealt{ng08}, \citealt{san09} and references therein). In Figure 12 we show the comparison between the
synthetic PL relations described here and those
corresponding to the chemical compositions of Cepheids in the
Magellanic Clouds (Z=0.004, Z=0.008 Y=0.25), and in the Milky Way
(Z=0.02 Y=0.28).  The PL relations for Z=0.0004 are steeper than those
at higher metallicities, confirming the trend with metallicity
predicted by our previous pulsation models \citep[see][]{b08,cmm00},
but are very similar to those at Z=0.004 for periods longer than about
10 days. The difference in absolute magnitude between the
predicted PL at Z=0.0004 and Z=0.004 is lower than 0.1 mag for
periods between about 8 and 70 days.

\section{Theory versus observations}

In this section we compare the theoretical relations derived in the
previous sections with the properties of the three Classical Cepheids
we have identified in IZw18 \citep{f10}.  It should be noted that the
PL is a statistical relation, and it is not safe to apply it to a very
small sample of Cepheids as that of IZw18.  
For this reason, in the following we will focus only on the Wesenheit
relations, that have much lower intrinsic scatter caused by the
finite width of the instability strip, and on the light curve model
fitting technique. In particular, due to the unusually long periods of
two of the 3 Cepheids found in IZw18 (namely, $P$=125 for V15, and
130.3 days for V1, see \citealt{f10},  for the identification of the
variables), we will apply these two theoretical methods only to the
bona fide classical Cepheid in IZw18, namely star V6 ($P$=8.71 days,
\citealt{f10}). 

In the two panels of Figure 13 we show the comparison between models
and observations in the $V,I$ Wesenheit versus period plane for two
different assumptions on the ML relation. As these relations are
reddening free, the shift required to superimpose the observed
distributions to the theoretical relations directly gives the
intrinsic distance modulus. By applying to V6 the canonical and
noncanonical Period-Wesenheit relations in the $V,I$ bands, we obtain
two estimates of the distance modulus of the galaxy and the associated
uncertainties, as labelled in each panel of Fig. 13. The two
inferred values are consistent with each other within the errors and in agreement with our previous estimates of the
distance to IZw18 obtained in \citet{a07}, and in particular with the
distance based on the TRGB ($\mu_0=31.30 \pm 0.17$ mag).

An independent constraint on the distance to IZw18 can also be
obtained by applying the light curve model fitting technique to V6.
This is a powerful tool to derive a direct estimate of the
intrinsic stellar parameters, that has already been applied with
success to both Galactic \citep{n08} and Magellanic
\citep{was97,bcm02,kw02,kw06} Cepheids. Results from the application
of the method to the observed $V, I$ light curves of V6 are shown in
Figs. 14 and 15, for isoperiodic sequences of models at fixed luminosity and at
fixed mass, respectively. These models reproduce the observed curves
within $\pm$ 0.13 mag simultaneously in $V$ and $I$. In turn, we find
that V6 is expected to have a stellar mass ranging from $5.9
M_{\odot}$ to $6.8 M_{\odot}$, $\log{L/L_{\odot}}$ in the range from
3.64 to 3.69 dex and $T_e$ varying from $\approx$ 5800 K to $\approx$
6000 K. The corresponding average apparent distance moduli in $V$ and
$I$ are $<\mu_V>=<\mu_I>= 31.5\pm 0.2$ mag, where the uncertainties
take into account both the observational error and the intrinsic
dispersion of the moduli obtained from the models shown in Figs. 14
and 15.  
By correcting the
obtained apparent moduli for the extinction in the two bands (adopting for
IZw18 a foreground Galactic extinction $E(B-V)$=0.032 mag, consistent with \citealt{a07} and \citealt{f10}), we find $\mu_0=31.4 \pm 0.2$ mag, again in
very good agreement with the distance modulus derived from the TRGB
method ($\mu_0=31.30 \pm 0.17$ mag).

As for the intrinsic stellar parameters inferred from the model
fitting technique, we note that the mass and luminosity estimated for
V6 are consistent with the evolutionary values obtained in
\citet{f10}. According to the canonical ML relation adopted in this
paper for a stellar luminosity $\log L/L_{\odot}=3.68 \pm 0.01$ dex, we
expect a stellar mass in the range from 6.0 to 6.5 $M_{\odot}$. Slightly
smaller masses (down to 5.8 $M_{\odot}$) can also been accepted, when
invoking a moderate overshooting ($\delta {\log{L/L_{\odot}}} <$0.25
dex). On the other hand,  stellar masses larger than $\sim$ 6.5
$M_{\odot}$ seem to be excluded for the quoted luminosity level, at
least on the basis of current stellar evolution models.  Taking into
account that the pulsation best fit model which minimizes the
residuals between predicted and observed light curves both in
$V$ and in $I$ bands, corresponds to a 6.5$M_{\odot}$ with $\log
L/L_{\odot}=3.68$ dex, that is very close to the canonical ML relation, we
may conclude that the light-curve model fitting of V6 favours the
canonical evolutionary scenario.

\section{Conclusions}

We updated our theoretical scenario for Classical Cepheids extending
it to the ultra-low metallicity composition of the blue compact
galaxy IZw18, i.e. Z=0.0004.  On the basis of these new models, we have
derived and listed the coefficients for the key Cepheid relations 
(namely PLC, Wesenheit and synthetic PL relations) at Z=0.0004.

In addition, we find that: 
\begin{enumerate}
\item The predicted IS topology confirms our previous
  results on its dependence on the ML relation and assumed
  mixing-length parameter.
\item The IS at Z=0.0004 is marginally bluer than that at Z=0.004.  
The small sensitivity to metallicity
  in this low Z regime leads to a leveling off
  of the metallicity effect on the synthetic PL relations.
\item The predicted bolometric light curves show the already known
  dependence on effective temperature and mixing-length parameter. For
  bump Cepheid models the transition from light curves with the bump
  on the decreasing branch to light curves with the bump on the
  increasing branch occurs at a pulsation period of about 13 days. This
  value is significantly longer than both observed \citep{m92,m00,be98}
  and predicted periods \citep{bms00} at the center of the HP ($P_{HP}$) 
  for Galactic, LMC and SMC bump Cepheids. This
  occurrence confirms the predicted and observed trend of $P_{HP}$ with
  metallicity \citep[see also][]{mmf05}.
\item The synthetic PL relations for Z=0.0004 are more linear and
  steeper than those obtained at higher Z, but very similar to the
  Z=0.004 ones, at periods longer than about 10 days.  The
  difference in absolute magnitude between the predicted PL at
  Z=0.0004 and Z=0.004 is lower than 0.1 mag for periods between
  about 8 and 70 days. This result indicates a significant reduction
  of the metallicity effect on the Cepheid PL relations for $ Z \le
  0.004$, thus supporting recent observational evidence of the
  universality of the PL slope in galaxies with metallicities
  significantly lower than in the LMC \citep[see e.g.][]{pie06,pie07}.
\item The application to the IZw18 Cepheid V6 of the Wesenheit relations 
 derived with both canonical and noncanonical assumptions,  
  provides two distance moduli consistent within
  the uncertainties with each other and
  with the distance to the galaxy inferred from the TRGB method.
\item Also the model fitting of the light curves of V6 provides 
 a distance modulus in good agreement with the TRGB distance.
\item The intrinsic stellar properties obtained from the model
  fitting of V6 are consistent with the evolutionary estimates based
  on current sets of stellar models slightly favouring the canonical scenario. 
\end{enumerate}

Finally, we notice that, further observations of Cepheids in very
metal poor galaxies are needed in order to clarify the nature of the
very long period candidates discovered in IZw18. If they were confirmed to
be the extension of Classical Cepheids to higher masses, they would provide
key constraints on pulsation models of massive stars.

\acknowledgements
MM, IM, GC and MT acknowledge financial support from PRIN-INAF2005, PRIN-INAF2006 and 
COFIS ASI-INAF I/016/07/0. Support for Francesca Annibali was also provided by NASA through
grants associated with program GO-10586.

\pagebreak

\bibliographystyle{apj}


\begin{table*}\label{strip}
\caption{Predicted location of the IS boundaries for fundamental mode pulsators with 
 Z=0.0004 and Y=0.24.}
\begin{center}
\begin{tabular}{lccc}
\hline \hline
$M/M_{\odot}$ & $log{L/L_{\odot}}$ & $T_e(FBE)$& $T_e(FRE)$ \\
4.0 &3.07 &6250 &5450\\
4.0 &3.32 &6150 &5250\\
5.0 &3.37 &6050 &5250\\
5.0 &3.62 & 6050  &5050\\
7.0 &3.84 & 5950 &4950\\
7.0 &4.09 & 5850 & 4650\\
9.0 &4.16 & 5850 &4750\\
9.0 &4.41 & 5850  &4650\\
11.0 & 4.42& 5750  &4750 \\
11.0 & 4.67& 5750  & 4900 \\
13.0 & 4.65 & 5650 & 4550\\
\hline
\end{tabular}
\end{center}
\end{table*}

\begin{table*}\label{stripFO}
\caption{Predicted location of the IS boundaries for first overtone pulsators with Z=0.0004 and Y=0.24.}
\begin{center}
\begin{tabular}{lccc}
\hline \hline
$M/M_{\odot}$ & $log{L/L_{\odot}}$ & $T_e(FOBE)$ & $T_e(FORE)$\\
4.0 &3.07 &6450 &6050 \\
4.0 &3.32 &6250 &5650 \\
4.5 &3.50 &6050 &5750 \\
5.0 &3.37 &6250 &6050 \\
7.0 &3.84 &5850 &5650 \\
\hline
\end{tabular}
\end{center}
\end{table*}

\begin{table}
\begin{center}
\caption{Intrinsic stellar parameters, periods and intensity-averaged mean magnitudes and colors for fundamental models with Z=0.0004, Y=0.24, and $l/H_p=1.5$.}
\begin{tabular}{lllllccccc}
\hline\hline 
{$M$}         &
{$logL$}      &
{$T_e$}       &
{$\log{P}$}   &
{$M_V$}       &
{$B-V$}       &
{$V-R$}       &
{$V-I$}       &
{$V-J$}       &
{$V-K$} \\ 
\hline
4.0 &3.07 &6200 & 0.4758 & $-$2.8917 &  0.3132 &  0.2372 &  0.5276 &  0.8947 &  1.2327 \\
4.0 &3.07 &6100 & 0.5001 & $-$2.8826 &  0.3465 &  0.2553 &  0.5615 &  0.9466 &  1.3010 \\
4.0 &3.07 &6000 & 0.5244 & $-$2.8713 &  0.3855 &  0.2751 &  0.5975 &  0.9998 &  1.3699 \\
4.0 &3.07 &5900 & 0.5494 & $-$2.8559 &  0.4273 &  0.2954 &  0.6343 &  1.0539 &  1.4405 \\
4.0 &3.07 &5800 & 0.5752 & $-$2.8389 &  0.4631 &  0.3142 &  0.6691 &  1.1060 &  1.5096 \\
4.0 &3.07 &5700 & 0.6010 & $-$2.8224 &  0.4935 &  0.3303 &  0.6992 &  1.1527 &  1.5735 \\
4.0 &3.07 &5600 & 0.6268 & $-$2.8071 &  0.5213 &  0.3446 &  0.7260 &  1.1951 &  1.6318 \\
4.0 &3.07 &5500 & 0.6529 & $-$2.7928 &  0.5472 &  0.3577 &  0.7504 &  1.2337 &  1.6847 \\
4.0 &3.32 &6100 & 0.7133 & $-$3.5142 &  0.3498 &  0.2550 &  0.5593 &  0.9389 &  1.2845 \\
4.0 &3.32 &6000 & 0.7386 & $-$3.4995 &  0.3887 &  0.2749 &  0.5959 &  0.9939 &  1.3579 \\
4.0 &3.32 &5900 & 0.7650 & $-$3.4831 &  0.4286 &  0.2952 &  0.6332 &  1.0497 &  1.4321 \\
4.0 &3.32 &5800 & 0.7916 & $-$3.4650 &  0.4644 &  0.3144 &  0.6691 &  1.1042 &  1.5053 \\
4.0 &3.32 &5700 & 0.8184 & $-$3.4467 &  0.4979 &  0.3321 &  0.7022 &  1.1561 &  1.5764 \\
4.0 &3.32 &5600 & 0.8460 & $-$3.4285 &  0.5289 &  0.3482 &  0.7328 &  1.2056 &  1.6455 \\
4.0 &3.32 &5500 & 0.8750 & $-$3.4122 &  0.5565 &  0.3623 &  0.7593 &  1.2484 &  1.7055 \\
4.0 &3.32 &5400 & 0.9050 & $-$3.3901 &  0.5908 &  0.3792 &  0.7919 &  1.3036 &  1.7860 \\
4.0 &3.32 &5300 & 0.9302 & $-$3.3703 &  0.6199 &  0.3931 &  0.8192 &  1.3508 &  1.8556 \\
5.0 &3.37 &5900 & 0.7413 & $-$3.6163 &  0.4183 &  0.2918 &  0.6283 &  1.0460 &  1.4316 \\
\hline
\end{tabular}
\end{center}
A portion of Table\,3 is shown here for guidance regarding its form
and content. The entire catalog is available in the electronic edition 
of the journal.
\end{table}
\clearpage
\begin{table*}\label{meanvalues}
\caption{As in Table 3, but for first overtone  models at Z=0.0004, Y=0.24 and $l/H_p=1.5$.}
\begin{center}
\begin{tabular}{lccccccccc}
\hline \hline
$M$ & $logL$ & $T_e$  &  $\log{P}$ & $M_V$ & $B-V$ &  $V-R$ &  $V-I$  &  $V-J$ & $V-K$ \\
4.0 &3.07 &6400 & 0.2975 & $-$2.9293 &  0.2826 &  0.2123 &  0.4747 &  0.7993 &  1.0811 \\
4.0 &3.07 &6300 & 0.3194 & $-$2.9170 &  0.3092 &  0.2282 &  0.5053 &  0.8478 &  1.1494 \\
4.0 &3.07 &6200 & 0.3424 & $-$2.9049 &  0.3440 &  0.2463 &  0.5389 &  0.8982 &  1.2171 \\
4.0 &3.07 &6100 & 0.3667 & $-$2.8907 &  0.3774 &  0.2644 &  0.5728 &  0.9492 &  1.2856 \\
4.0 &3.32 &6200 & 0.5470 & $-$3.5317 &  0.3553 &  0.2504 &  0.5448 &  0.9005 &  1.2123 \\
4.0 &3.32 &6100 & 0.5712 & $-$3.5150 &  0.3670 &  0.2614 &  0.5688 &  0.9473 &  1.2879 \\
4.0 &3.32 &6000 & 0.5971 & $-$3.4979 &  0.4000 &  0.2798 &  0.6038 &  1.0025 &  1.3651 \\
4.0 &3.32 &5900 & 0.6203 & $-$3.4808 &  0.4359 &  0.2988 &  0.6394 &  1.0573 &  1.4399 \\
4.0 &3.32 &5800 & 0.6449 & $-$3.4629 &  0.4685 &  0.3166 &  0.6730 &  1.1097 &  1.5121 \\
4.0 &3.32 &5700 & 0.6695 & $-$3.4453 &  0.4992 &  0.3330 &  0.7038 &  1.1590 &  1.5807 \\
4.5 &3.50 &6000 & 0.7110 & $-$3.9494 &  0.4022 &  0.2801 &  0.6038 &  1.0010 &  1.3611 \\
4.5 &3.50 &5800 & 0.7585 & $-$3.9127 &  0.4683 &  0.3167 &  0.6733 &  1.1109 &  1.5147 \\
5.0 &3.37 &6200 & 0.5305 & $-$3.6643 &  0.3470 &  0.2466 &  0.5385 &  0.8953 &  1.2104 \\
5.0 &3.37 &6100 & 0.5538 & $-$3.6490 &  0.3732 &  0.2625 &  0.5697 &  0.9463 &  1.2837 \\
7.0 &3.84 &5800 & 0.9198 & $-$4.7600 &  0.4640 &  0.3148 &  0.6703 &  1.1083 &  1.5134 \\
7.0 &3.84 &5700 & 0.9439 & $-$4.7437 &  0.4986 &  0.3331 &  0.7048 &  1.1628 &  1.5889 \\
\hline
\end{tabular}
\end{center}
\end{table*}

\begin{table*}\label{plc}
\caption{Coefficients of the multifilter  PLC relations for fundamental mode models 
with Z=0.0004, Y=0.24, as a function of the adopted  ML relation, in the form  
$<M_V>$=$\alpha$+$\beta$log$P$+$\gamma$(color)+$\delta${$\log{L/L_C}$}}
\begin{center}
\begin{tabular}{lcccccccccc}
\hline \hline
$color$ & $\alpha$ & $\beta$ & $\gamma$ & $\delta$ & $\sigma$ \\
\hline
B-V & $-$2.28 & $-$3.78 & 3.47 & 0.90 & 0.03\\
& $\pm$0.01&$\pm$0.01&$\pm$0.04& $\pm$0.03 &\\
V-R & $-$3.03    &  $-$3.79       & 7.50  & 0.92     &    0.05     \\
& $\pm$0.05&$\pm$0.02&$\pm$0.12&$\pm$0.04   &\\
V-I & $-$3.35 & $-$3.81 & 4.00 & 0.93 & 0.05  \\
& $\pm$0.05&$\pm$0.02&$\pm$0.06&$\pm$0.04 &\\
V-J & $-$3.42 & $-$3.82 & 2.49 & 0.95 &0.04 \\
& $\pm$0.04&$\pm$0.02&$\pm$0.04&$\pm$0.04 &\\
V-K & $-$3.36 & $-$3.84 & 1.79 & 0.97 & 0.05\\
& $\pm$0.05&$\pm$0.02&$\pm$0.03&$\pm$0.04 &\\
\hline
\end{tabular}
\end{center}
\end{table*}

\begin{table*}\label{wes}
\caption{Coefficients of the multi-filter Wesenheit relations for fundamental mode 
models with Z=0.0004, Y=0.24, as a function of the adopted  ML relation, in the form  
$<M_V> - \gamma$(color)=$\alpha_W$+$\beta_W$log$P$+$\delta_W${$\log{L/L_C}$}}
\begin{center}
\begin{tabular}{lccccccccc}
\hline \hline
Wesenheit function & $\alpha_W$ & $\beta_W$ & $\delta_W$ & $\sigma$ \\
\hline
$<M_V> -3.10 (<B>-<V>)$ & $-$2.17 & $-$3.69 & 0.88 & 0.05\\
& $\pm0.05$&$\pm$0.01&$\pm$0.04&  \\
$<M_V> -6.29 (<V>-<R>)$ & $-$2.74 & $-$3.66 & 0.88 & 0.07 \\
& $\pm0.07$&$\pm$0.02&$\pm$0.06&  \\
$<M_V> -2.54 (<V>-<I>)$ & $-$2.58 & $-$3.49 & 0.84 & 0.12 \\
& $\pm$0.12 & $\pm$0.03 & $\pm$0.10& \\ 
$<M_V> -1.39 (<V>-<J>)$ & $-$2.46 & $-$3.43 & 0.84 & 0.14 \\
& $\pm$0.14 & $\pm$0.03 & $\pm$0.12& \\ 
$<M_V> -1.13 (<V>-<K>)$ & $-$2.58 & $-$3.51 & 0.87 & 0.12 \\
& $\pm$0.12 & $\pm$0.03 & $\pm$0.10& \\ 
\hline
\end{tabular}
\end{center}
\end{table*}

\begin{table*}\label{plcan}
\caption{Coefficients of the linear synthetic PL relations based on canonical 
fundamental mode models with Z=0.0004, Y=0.24 and $l/H_P=1.5$. Relations have the 
form $M_j=A_c + B_c  \log P$, where j is the photometric band.}
\begin{center}
\begin{tabular}{lcccccccc}
\hline \hline
Magnitude & $A_c$ & $B_c$ & $\sigma$ \\
\hline
$M_B$ & $-$0.76 & $-$2.81 & 0.28\\
 &$\pm$0.02&$\pm$0.02&  \\
$M_V$ & $-$1.10 & $-$3.04 & 0.22 \\
 &$\pm$0.02&$\pm$0.02&  \\
$M_R$ & $-$1.36 & $-$3.14 & 0.18 \\
 & $\pm$0.01 & $\pm$0.01& \\ 
$M_I$ & $-$1.67 & $-$3.24 & 0.16 \\
 & $\pm$0.01 & $\pm$0.01& \\ 
$M_J$ & $-$2.03 & $-$3.37 & 0.12 \\
 & $\pm$0.01 & $\pm$0.01& \\ 
$M_K$ & $-$2.36 & $-$3.50 & 0.07 \\
 & $\pm$0.01 & $\pm$0.01& \\ 
\hline
\end{tabular}
\end{center}
\end{table*}

\begin{table*}\label{plalpha}
\caption{Coefficients of the linear synthetic PL relations based on canonical 
fundamental mode models with Z=0.0004, Y=0.24 and $l/H_P=2.0$. Relations have the 
form $M_j=A_c + B_c \log P$, where j is the photometric band.}
\begin{center}
\begin{tabular}{lcccccccc}
\hline \hline
Magnitude & $A_c$ & $B_c$ & $\sigma$ \\
\hline
$M_B$ & $-$0.97&$-$2.80  & 0.19\\
 &$\pm$0.02&$\pm$0.02&  \\
$M_V$ & $-$1.27 & $-$3.03 & 0.16 \\
 &$\pm$0.02&$\pm$0.02&  \\
$M_R$ & $-$1.51 & $-$3.13 & 0.15 \\
 & $\pm$0.02 & $\pm$0.02& \\ 
$M_I$ & $-$1.79 & $-$3.23 & 0.14 \\
 & $\pm$0.01 & $\pm$0.01& \\ 
$M_J$ & $-$2.12 & $-$3.36 & 0.12 \\
 & $\pm$0.01 & $\pm$0.01& \\ 
$M_K$ & $-$2.43 & $-$3.48 & 0.10 \\
 & $\pm$0.01 & $\pm$0.01& \\ 
\hline
\end{tabular}
\end{center}
\end{table*}

\begin{table*}\label{plnoncan}
\caption{Coefficients of the linear synthetic PL relations based on noncanonical 
fundamental mode models with Z=0.0004, Y=0.24 and $l/H_P=1.5$. Relations have the 
form $M_j=A_{nc} + B_{nc} \log P$, where j is the photometric band.}
\begin{center}
\begin{tabular}{lcccccccc}
\hline \hline
Magnitude & $A_{nc}$ & $B_{nc}$ & $\sigma$ \\
\hline
$M_B$ &$-$1.01 & $-$2.54 & 0.36 \\
 &$\pm$0.04&$\pm$0.04&  \\
$M_V$ & $-$1.22 & $-$2.85 & 0.27 \\
 &$\pm$0.02&$\pm$0.02&  \\
$M_R$ & $-$1.42 & $-$2.98 & 0.23 \\
 & $\pm$0.02 & $\pm$0.02& \\ 
$M_I$ & $-$1.68 & $-$3.10 & 0.20 \\
 & $\pm$0.02 & $\pm$0.02& \\ 
$M_J$ & $-$1.98 & $-$3.25 & 0.14 \\
 & $\pm$0.02 & $\pm$0.02& \\ 
$M_K$ & $-$2.26 & $-$3.40 & 0.09 \\
 & $\pm$0.01 & $\pm$0.01& \\ 
\hline
\end{tabular}
\end{center}
\end{table*}

\clearpage

\begin{figure} \begin{center}
\includegraphics[width=15cm]{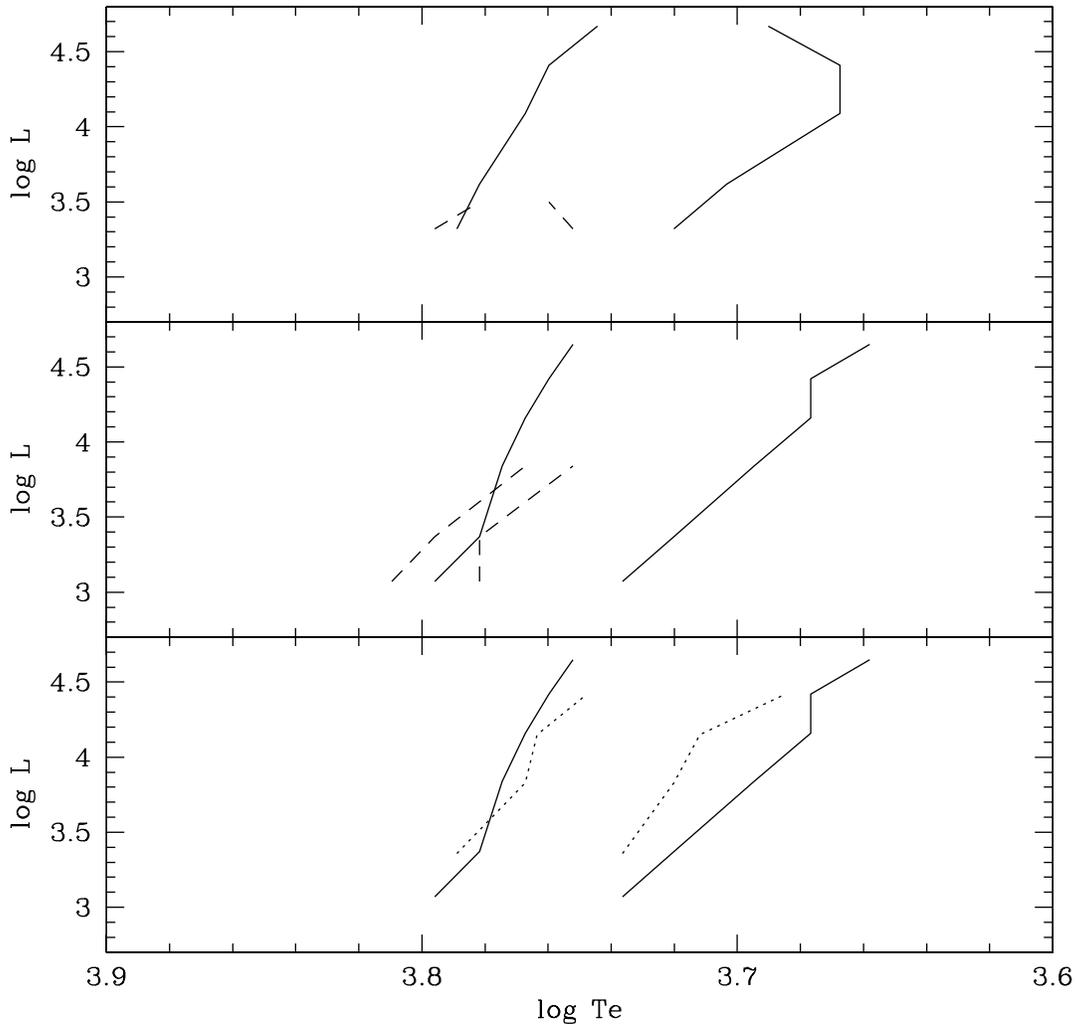}
\caption[]{Location in the HR diagram of the fundamental (solid lines)
  and first overtone (dashed lines) IS boundaries for
  canonical (middle panel) and noncanonical (upper panel) models with
  $l/H_p=1.5$. In the lower panel the theoretical fundamental boundaries
  for canonical models with $l/H_p=1.5$ (solid lines) are compared
  with the same predictions at $l/H_p=2.0$ (dotted lines).}
\label{f1}
\end{center} \end{figure}

\clearpage

\begin{figure} \begin{center}
\vspace{-3 cm}
\includegraphics[width=15cm]{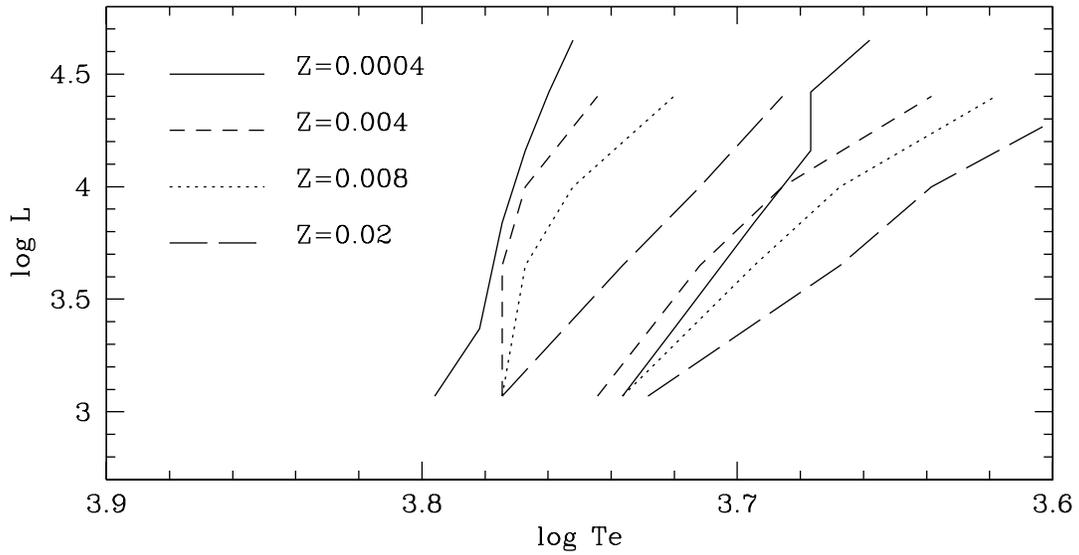}
\caption[]{Comparison between the theroretical IS obtained
  in the present paper for fundamental canonical models with
  $l/H_p=1.5$ and results obtained by \citet{bms99} for $Z=0.004$,
  $Z=0.008$ and $Z=0.02$.}
\label{f2}
\end{center} \end{figure}

\clearpage

\begin{figure} \begin{center}
\includegraphics[width=15cm]{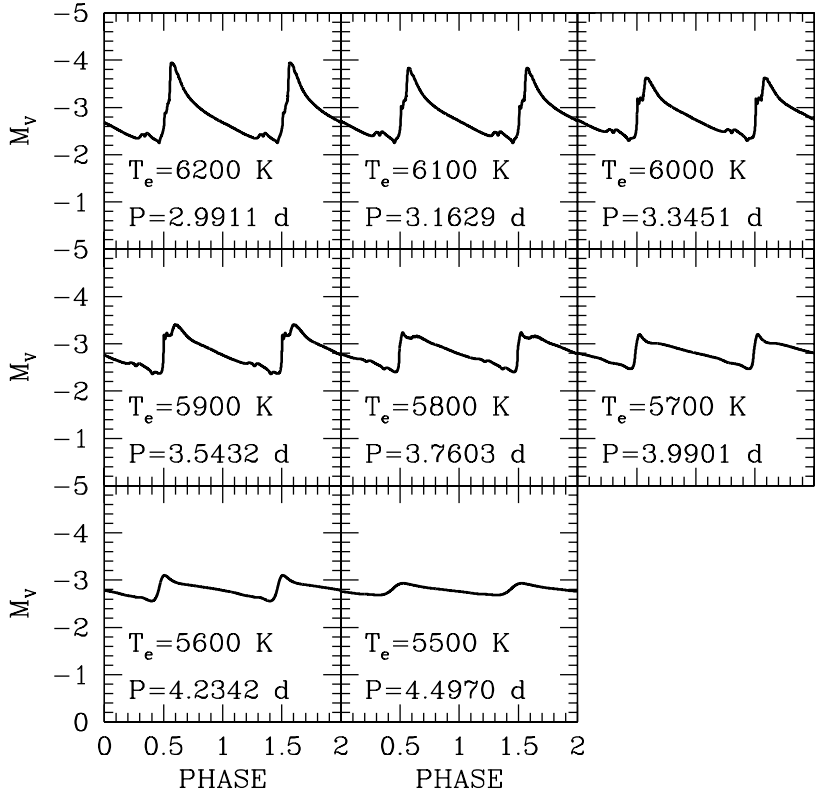}
\vspace{-2 cm}
\caption[]{Predicted bolometric light curves for canonical fundamental
  pulsation models with $M=4M_{\odot}$.}
\label{f3}
\end{center} \end{figure}

\clearpage

\begin{figure} \begin{center}
\includegraphics[width=15cm]{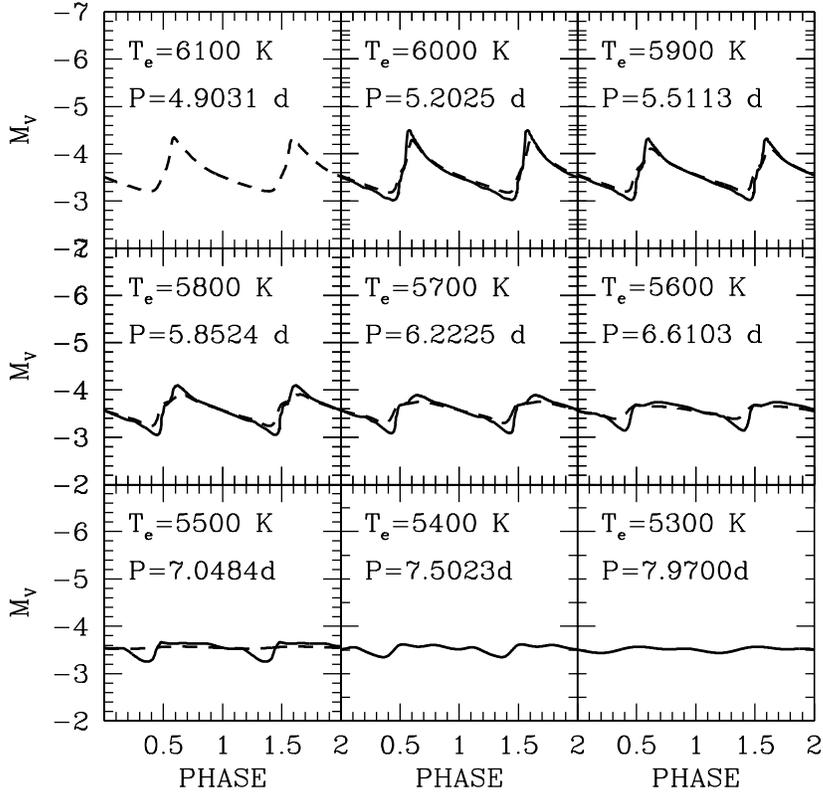}
\vspace{-2 cm}
\caption[]{Same as Figure 3, but for $M=5M_{\odot}$. The dashed lines are
  the corresponding curves for $l/H_p=2.0$.}
\label{f4}
\end{center} \end{figure}

\clearpage

\begin{figure} \begin{center}
\includegraphics[width=15cm]{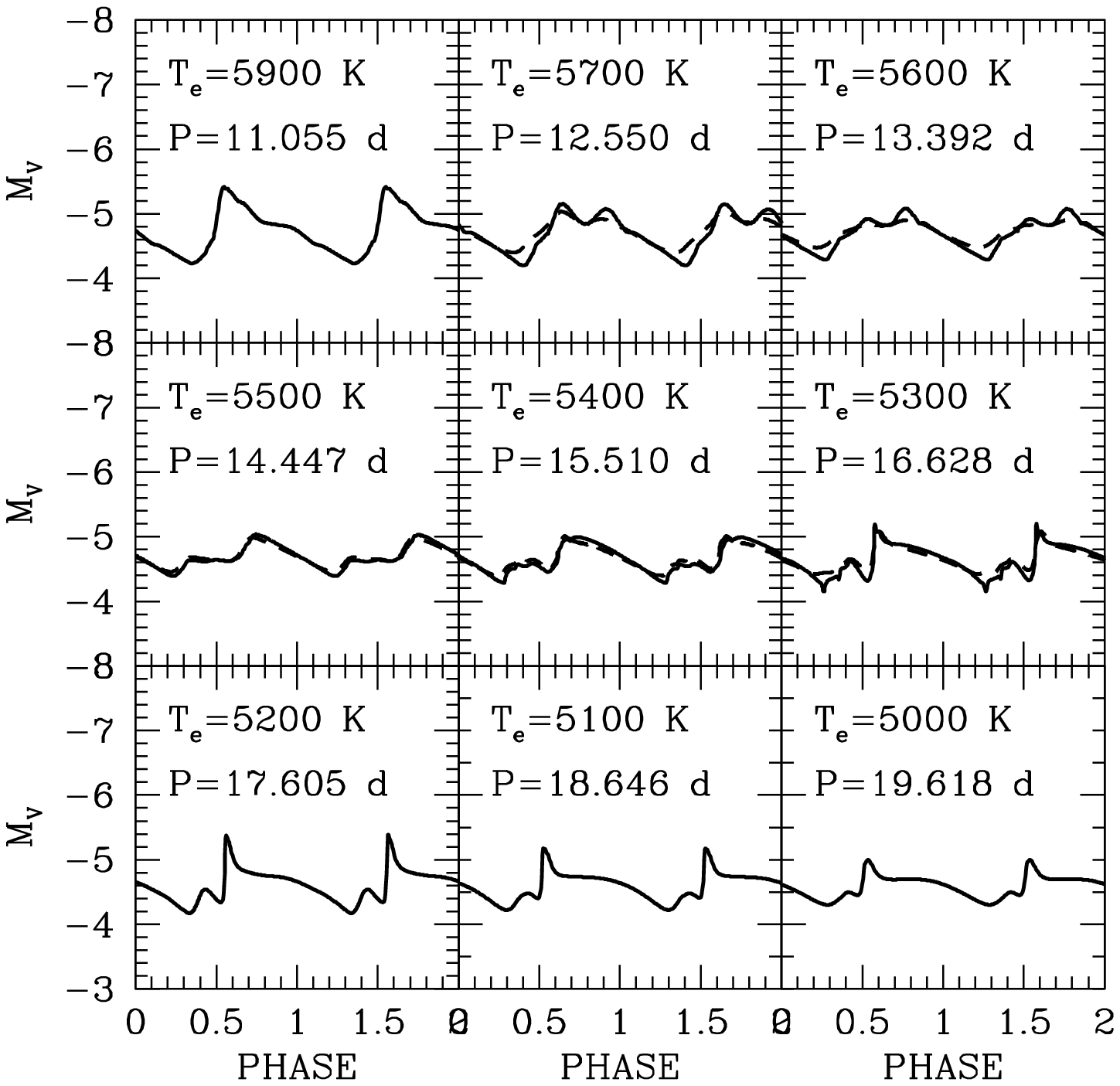}
\vspace{-2 cm}
\caption[]{Same as Figure 4, but for $M=7M_{\odot}$.}
\label{f5}
\end{center} \end{figure}

\clearpage

\begin{figure} \begin{center}
\includegraphics[width=15cm]{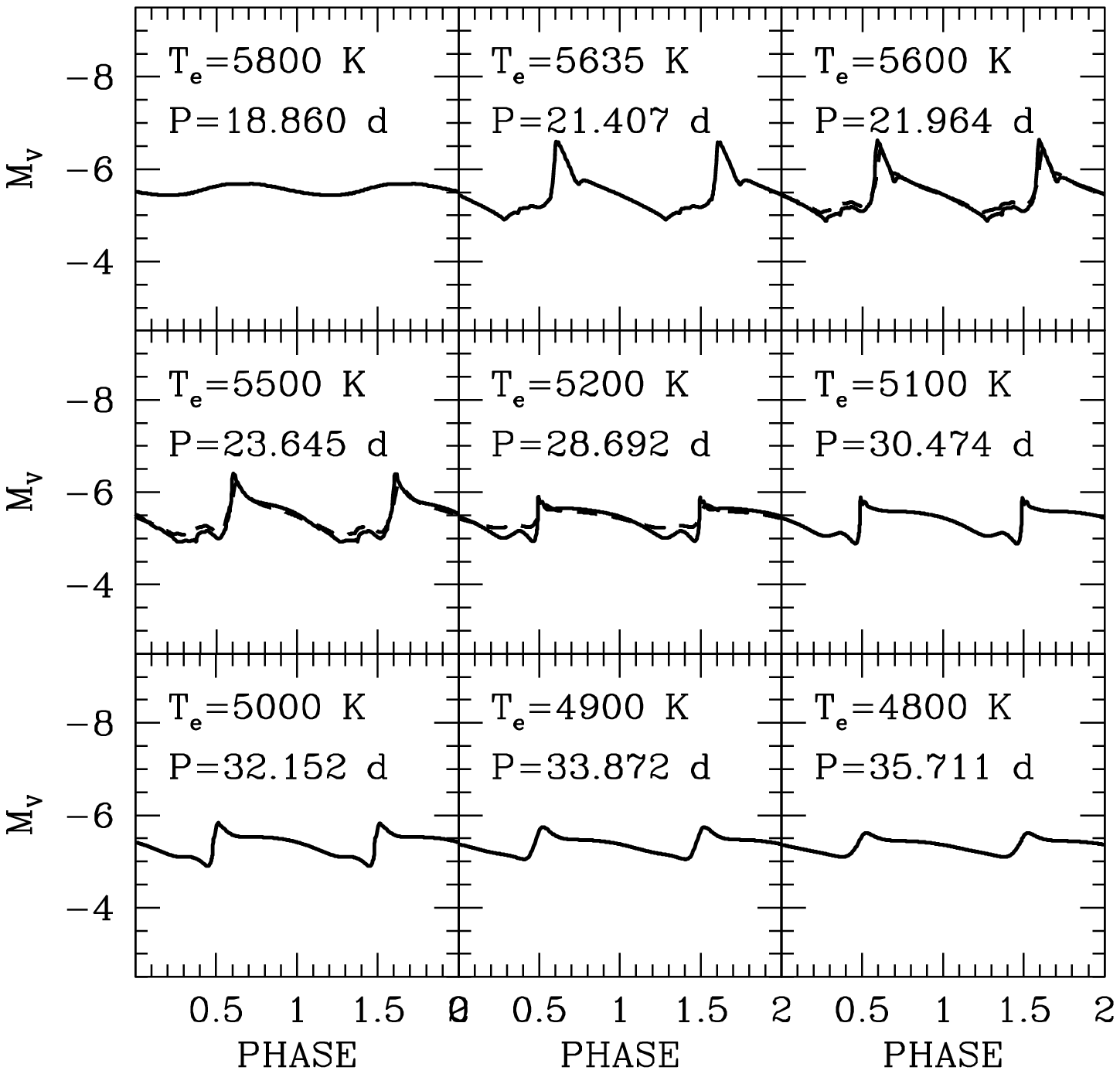}
\vspace{-2 cm}
\caption[]{Same as Figure 4, but for $M=9M_{\odot}$.} 
\label{f6}
\end{center} \end{figure}

\clearpage

\begin{figure} \begin{center}
\includegraphics[width=15cm]{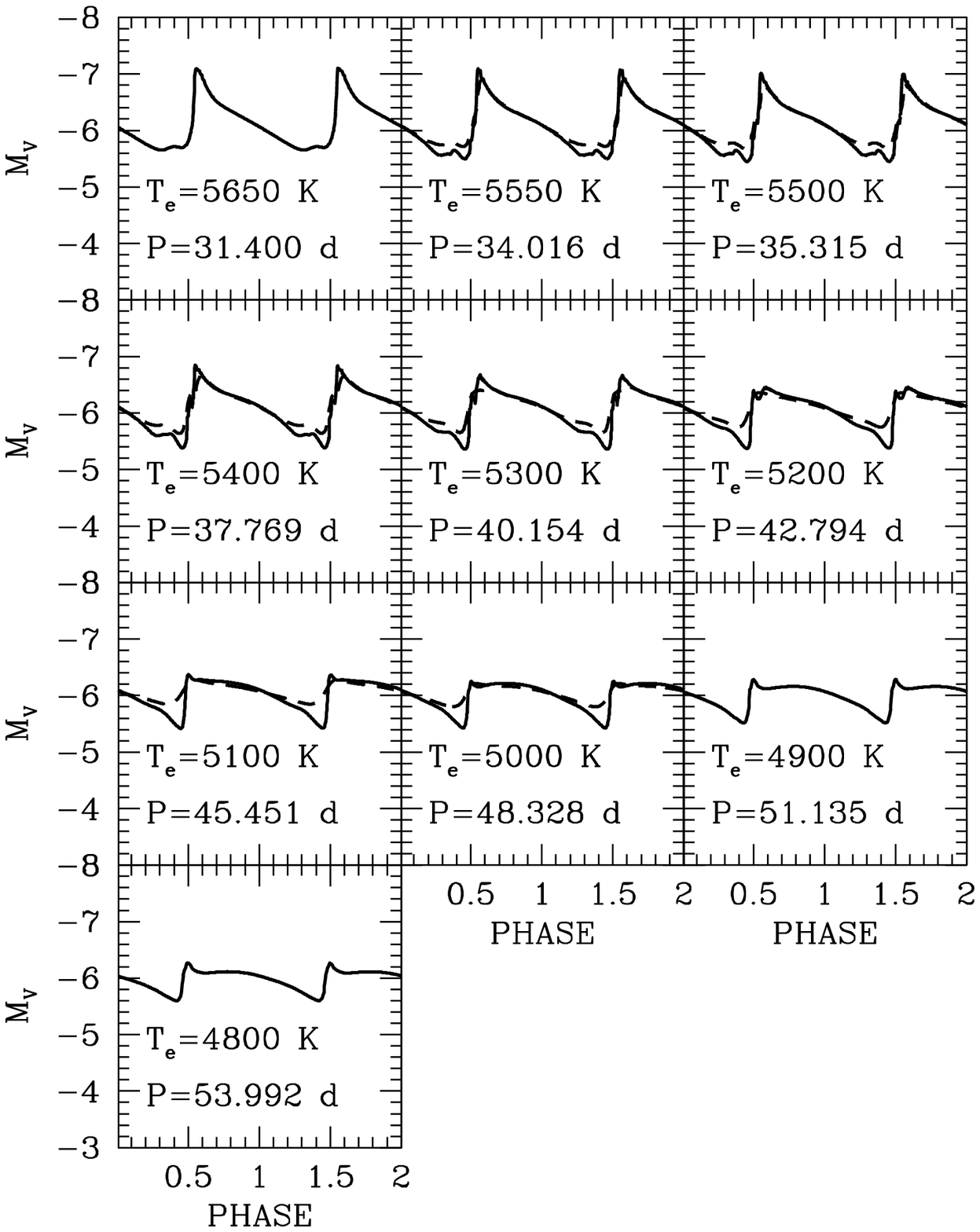}
\caption[]{Same as Figure 4, but for $M=11M_{\odot}$.} 
\label{f7}
\end{center} \end{figure}

\clearpage

\begin{figure} \begin{center}
\includegraphics[width=15cm]{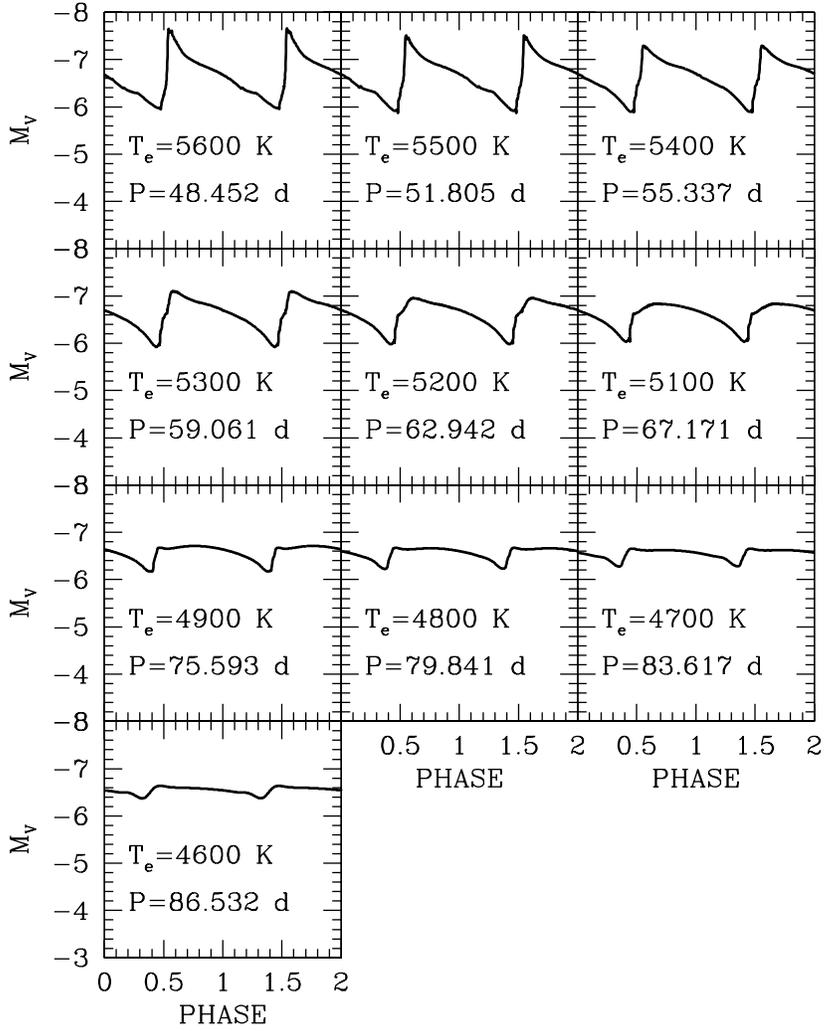}
\caption[]{Same as Figure 3, but for $M=13M_{\odot}$.} 
\label{f8}
\end{center} \end{figure}

\clearpage

\begin{figure} \begin{center}
\includegraphics[width=15cm]{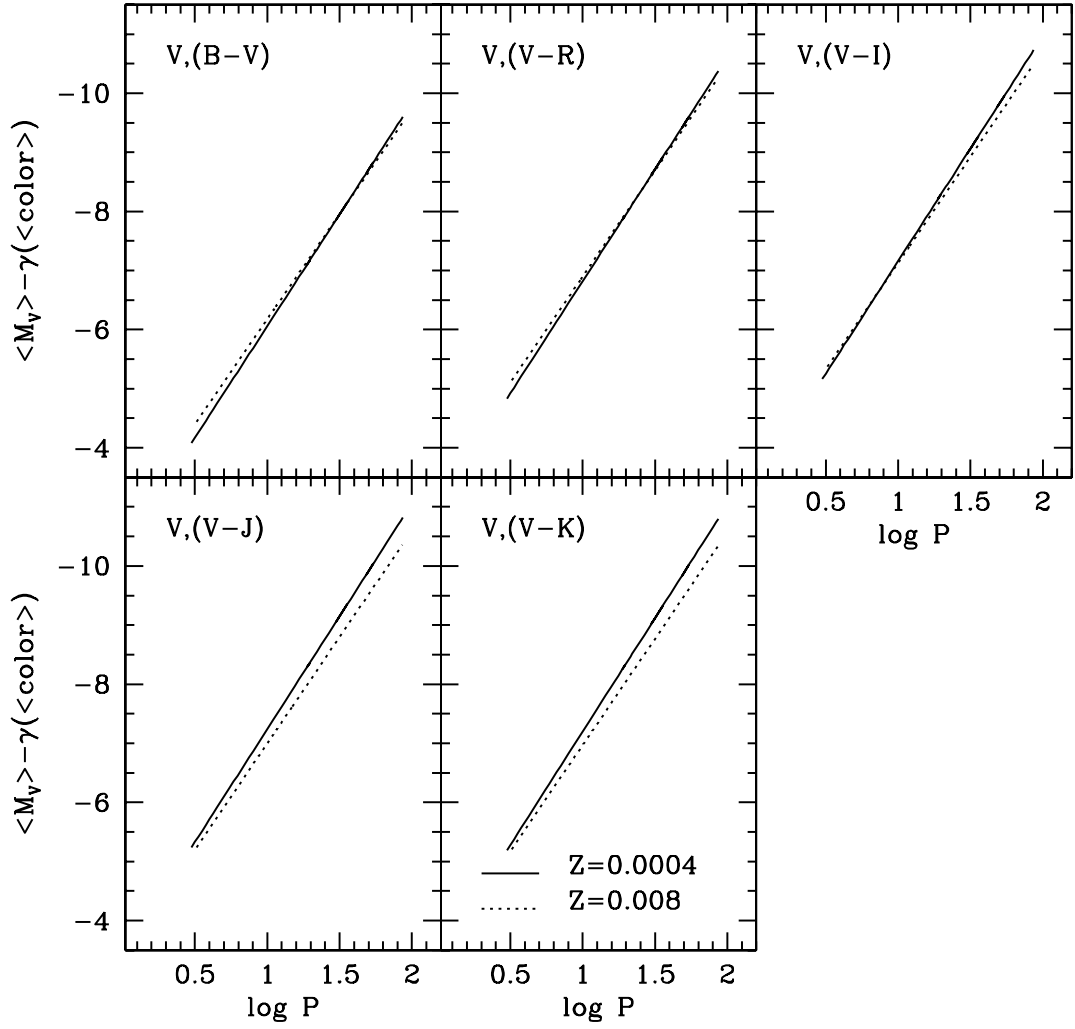}
\caption[]{Comparison between theoretical PLC relations at Z=0.0004
  and Z=0.008.}
\label{f9}
\end{center} \end{figure}

\clearpage

\begin{figure} \begin{center}
\includegraphics[width=15cm]{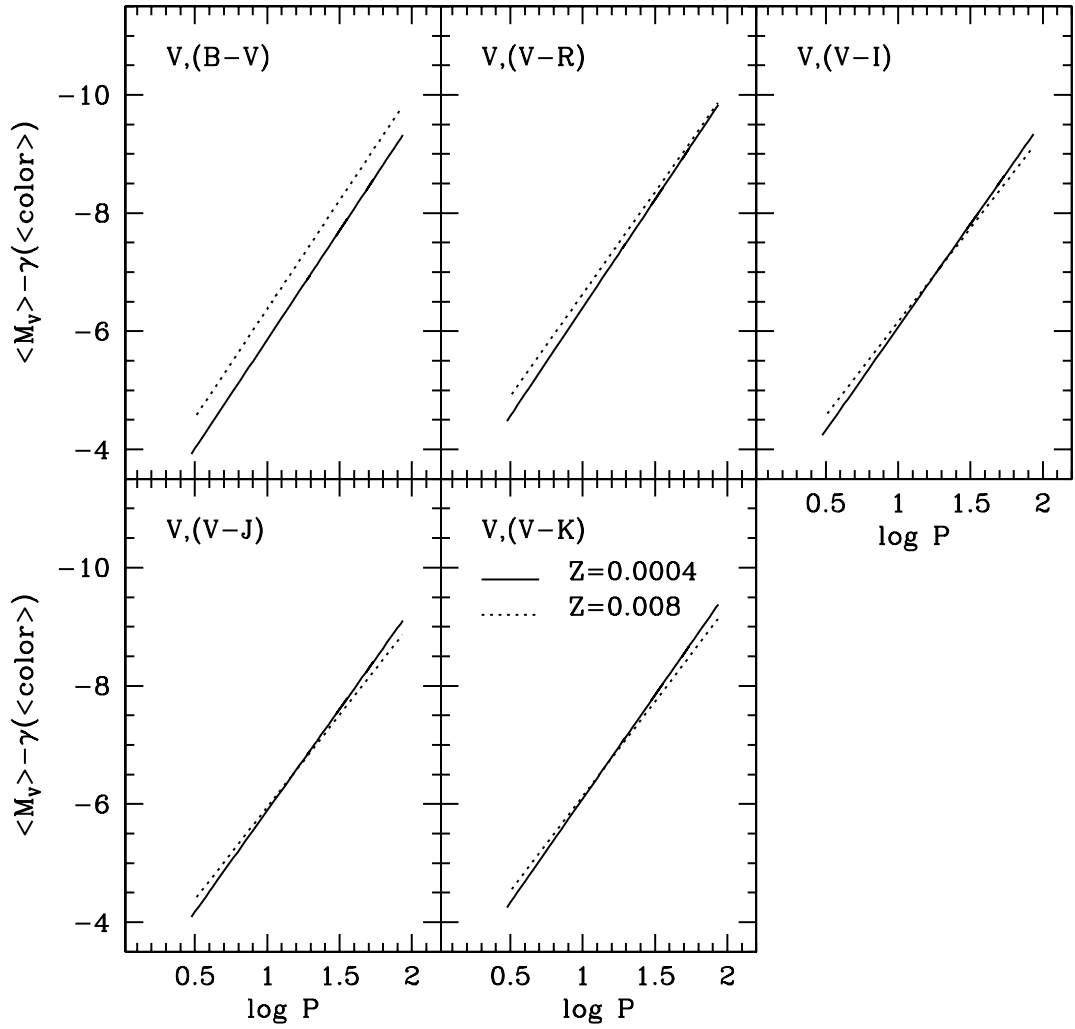}
\caption[]{Comparison between theoretical Period-Wesenheit relations
  at Z=0.0004 and Z=0.008.}
\label{f10}
\end{center} \end{figure}

\clearpage

\begin{figure} \begin{center}
\caption[]{Synthetic multi-filter PL relations for canonical
  fundamental pulsators and two different assumptions on $l/H_p$
  (left and right panels), and for noncanonical models with $l/H_p=1.5$
  (middle panel).}
\label{f11}
\end{center} \end{figure}

\clearpage

\begin{figure} \begin{center}
\includegraphics[width=15cm]{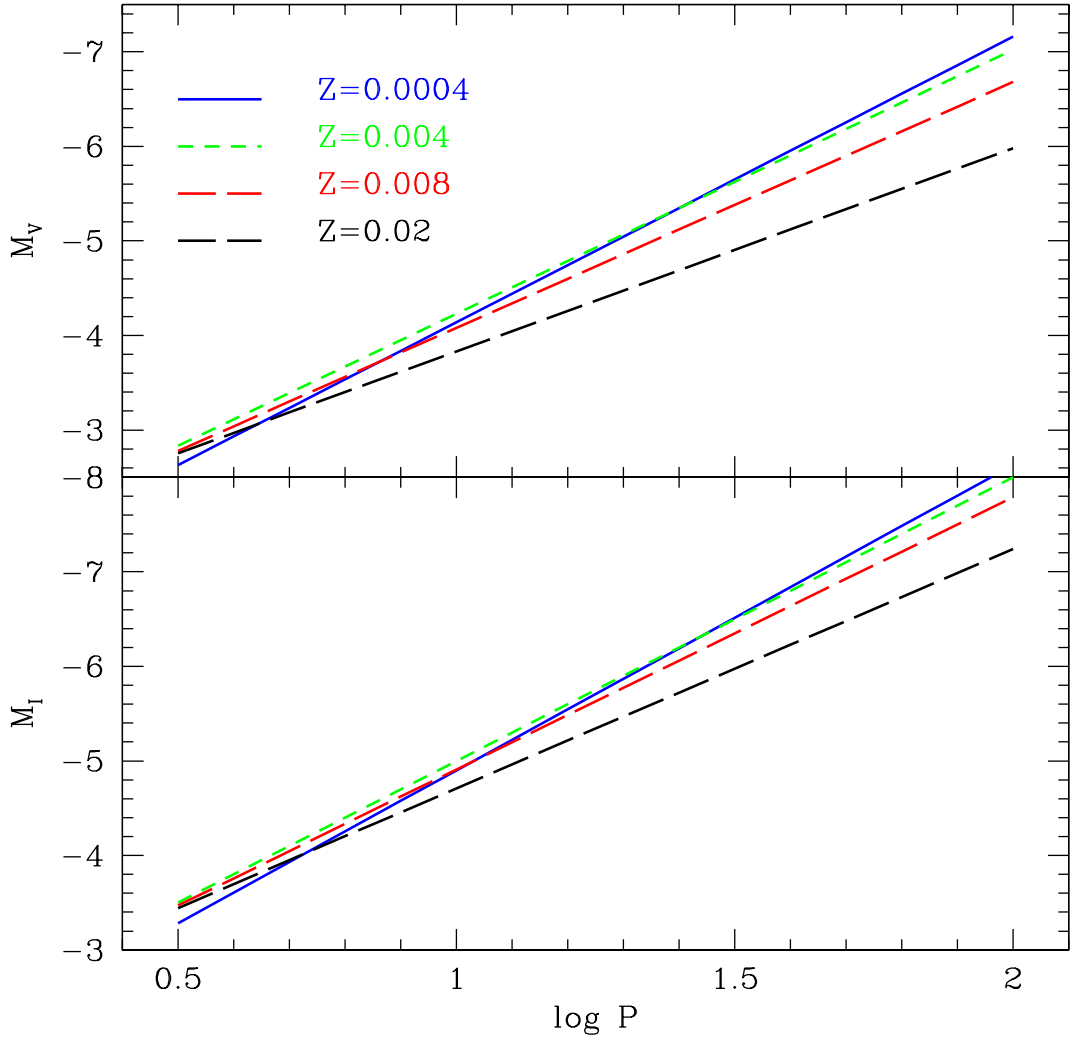}
\caption[]{Comparison between the synthetic PL relations obtained in
  this paper, and the relations corresponding to the chemical
  composition of Cepheids in the Magellanic Clouds, and in the Milky
  Way.}
\label{f12}
\end{center} \end{figure}

\clearpage

\begin{figure} \begin{center}
\includegraphics[width=15cm]{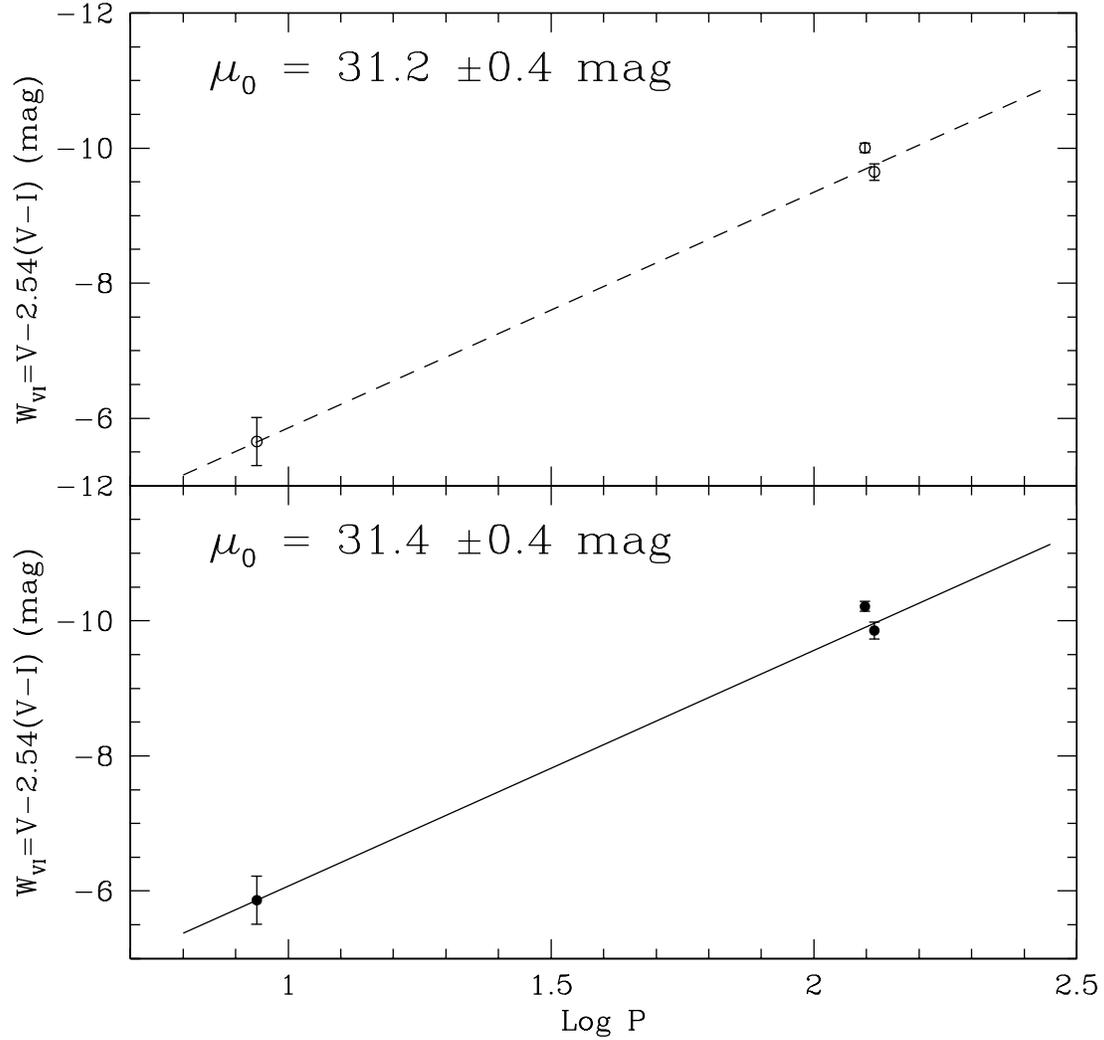}
\caption[]{Comparison between theoretical canonical (lower panel)
  and noncanonical (upper panel) Period-Wesenheit relations in the
  $VI$ bands and data of the three Cepheids in IZw18
  \citep[see][]{f10}. Distance moduli obtained from the
  application of these relations to the bona fide Classical Cepheid V6
  \citep[][]{f10} are labelled.}
\label{f13}
\end{center} \end{figure}

\clearpage

\begin{figure} \begin{center}
\vspace{-3 cm}
\vspace{-3 cm}
\caption[]{Model fitting of the light curves of V6 in the $V$ and $I$ band
  for a range of model masses and effective temperature (see figure's labels) at fixed
  luminosity level (see text).  All the plotted models are able to
  reproduce the observed $V,I$ light curves of V6 within 0.13 mag
  (mean residual).}
\label{f14}
\end{center} \end{figure}

\clearpage

\begin{figure} \begin{center}
\vspace{-4 cm}
\includegraphics[width=15cm]{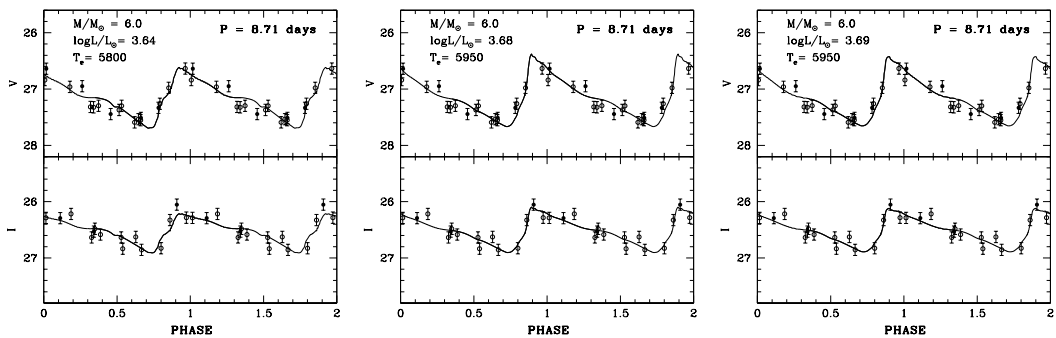}
\vspace{-4 cm}
\caption[]{Same as Fig. 14, but for models at fixed stellar mass.} 
\label{f15}
\end{center} \end{figure}

\end{document}